\documentstyle[12pt]{article}

\catcode`\@=11
\long\def\@makefntext#1{ 
\protect\noindent \hbox to 3.2pt {\hskip-.9pt

$^{{\ninerm\@thefnmark}}$\hfil}#1\hfill} 

\def\thefootnote{\fnsymbol{footnote}}
 \def\@makefnmark{\hbox to 0pt{$^{\@thefnmark}$\hss}}  

\def\ps@myheadings{\let\@mkboth\@gobbletwo
\def\@oddhead{\hbox{} 
\rightmark\hfil\ninerm\thepage}

\def\@oddfoot{}\def\@evenhead{\ninerm\thepage\hfil 
\leftmark\hbox{}}\def\@evenfoot{}
\def\sectionmark##1{}\def\subsectionmark##1{}}

\begin{document}

\newcommand{\symbolfootnote}{\renewcommand{\thefootnote}
	{\fnsymbol{footnote}}}
\renewcommand{\thefootnote}{\fnsymbol{footnote}}
\newcommand{\alphfootnote}
	{\setcounter{footnote}{0}
	 \renewcommand{\thefootnote}{\sevenrm\alph{footnote}}}

\newcounter{sectionc}\newcounter{subsectionc}\newcounter{subsubsectionc}
\renewcommand{\section}[1] {\vspace{0.6cm}\addtocounter{sectionc}{1}
\setcounter{subsectionc}{0}\setcounter{subsubsectionc}{0}\noindent
	{\bf\thesectionc. #1}\par\vspace{0.4cm}}
\renewcommand{\subsection}[1] {\vspace{0.6cm}\addtocounter{subsectionc}{1}
	\setcounter{subsubsectionc}{0}\noindent
	{\it\thesectionc.\thesubsectionc. #1}\par\vspace{0.4cm}}
\renewcommand{\subsubsection}[1]
{\vspace{0.6cm}\addtocounter{subsubsectionc}{1}
	\noindent {\rm\thesectionc.\thesubsectionc.\thesubsubsectionc.
	#1}\par\vspace{0.4cm}}
\newcommand{\nonumsection}[1] {\vspace{0.6cm}\noindent{\bf #1}
	\par\vspace{0.4cm}}

\newcounter{appendixc}
\newcounter{subappendixc}[appendixc]
\newcounter{subsubappendixc}[subappendixc]
\renewcommand{\thesubappendixc}{\Alph{appendixc}.\arabic{subappendixc}}
\renewcommand{\thesubsubappendixc}
	{\Alph{appendixc}.\arabic{subappendixc}.\arabic{subsubappendixc}}

\renewcommand{\appendix}[1] {\vspace{0.6cm}
        \refstepcounter{appendixc}
        \setcounter{figure}{0}
        \setcounter{table}{0}
        \setcounter{equation}{0}
        \renewcommand{\thefigure}{\Alph{appendixc}.\arabic{figure}}
        \renewcommand{\thetable}{\Alph{appendixc}.\arabic{table}}
        \renewcommand{\theappendixc}{\Alph{appendixc}}
        \renewcommand{\theequation}{\Alph{appendixc}.\arabic{equation}}
        \noindent{\bf Appendix \theappendixc #1}\par\vspace{0.4cm}}
\newcommand{\subappendix}[1] {\vspace{0.6cm}
        \refstepcounter{subappendixc}
        \noindent{\bf Appendix \thesubappendixc. #1}\par\vspace{0.4cm}}
\newcommand{\subsubappendix}[1] {\vspace{0.6cm}
        \refstepcounter{subsubappendixc}
        \noindent{\it Appendix \thesubsubappendixc. #1}
	\par\vspace{0.4cm}}

\def\abstracts#1{{
	\centering{\begin{minipage}{30pc}\tenrm\baselineskip=12pt\noindent
	\centerline{\tenrm ABSTRACT}\vspace{0.3cm}
	\parindent=0pt #1
	\end{minipage} }\par}}

\newcommand{\bibit}{\it}
\newcommand{\bibbf}{\bf}
\renewenvironment{thebibliography}[1]
	{\begin{list}{\arabic{enumi}.}
	{\usecounter{enumi}\setlength{\parsep}{0pt}
\setlength{\leftmargin 1.25cm}{\rightmargin 0pt}
	 \setlength{\itemsep}{0pt} \settowidth
	{\labelwidth}{#1.}\sloppy}}{\end{list}}

\topsep=0in\parsep=0in\itemsep=0in
\parindent=1.5pc

\newcounter{itemlistc}
\newcounter{romanlistc}
\newcounter{alphlistc}
\newcounter{arabiclistc}
\newenvironment{itemlist}
    	{\setcounter{itemlistc}{0}
	 \begin{list}{$\bullet$}
	{\usecounter{itemlistc}
	 \setlength{\parsep}{0pt}
	 \setlength{\itemsep}{0pt}}}{\end{list}}

\newenvironment{romanlist}
	{\setcounter{romanlistc}{0}
	 \begin{list}{$($\roman{romanlistc}$)$}
	{\usecounter{romanlistc}
	 \setlength{\parsep}{0pt}
	 \setlength{\itemsep}{0pt}}}{\end{list}}

\newenvironment{alphlist}
	{\setcounter{alphlistc}{0}
	 \begin{list}{$($\alph{alphlistc}$)$}
	{\usecounter{alphlistc}
	 \setlength{\parsep}{0pt}
	 \setlength{\itemsep}{0pt}}}{\end{list}}

\newenvironment{arabiclist}
	{\setcounter{arabiclistc}{0}
	 \begin{list}{\arabic{arabiclistc}}
	{\usecounter{arabiclistc}
	 \setlength{\parsep}{0pt}
	 \setlength{\itemsep}{0pt}}}{\end{list}}

\newcommand{\fcaption}[1]{
        \refstepcounter{figure}
        \setbox\@tempboxa = \hbox{\tenrm Fig.~\thefigure. #1}
        \ifdim \wd\@tempboxa > 6in
           {\begin{center}
        \parbox{6in}{\tenrm\baselineskip=12pt Fig.~\thefigure. #1 }
            \end{center}}
        \else
             {\begin{center}
             {\tenrm Fig.~\thefigure. #1}
              \end{center}}
        \fi}

\newcommand{\tcaption}[1]{
        \refstepcounter{table}
        \setbox\@tempboxa = \hbox{\tenrm Table~\thetable. #1}
        \ifdim \wd\@tempboxa > 6in
           {\begin{center}
        \parbox{6in}{\tenrm\baselineskip=12pt Table~\thetable. #1 }
            \end{center}}
        \else
             {\begin{center}
             {\tenrm Table~\thetable. #1}
              \end{center}}
        \fi}

\def\@citex[#1]#2{\if@filesw\immediate\write\@auxout
	{\string\citation{#2}}\fi
\def\@citea{}\@cite{\@for\@citeb:=#2\do
	{\@citea\def\@citea{,}\@ifundefined
	{b@\@citeb}{{\bf ?}\@warning
	{Citation `\@citeb' on page \thepage \space undefined}}
	{\csname b@\@citeb\endcsname}}}{#1}}

\newif\if@cghi
\def\cite{\@cghitrue\@ifnextchar [{\@tempswatrue
	\@citex}{\@tempswafalse\@citex[]}}
\def\citelow{\@cghifalse\@ifnextchar [{\@tempswatrue
	\@citex}{\@tempswafalse\@citex[]}}
\def\@cite#1#2{{$\null^{#1}$\if@tempswa\typeout
	{IJCGA warning: optional citation argument
	ignored: `#2'} \fi}}
\newcommand{\citeup}{\cite}

\def\fnm#1{$^{\mbox{\scriptsize #1}}$}
\def\fnt#1#2{\footnotetext{\kern-.3em
	{$^{\mbox{\sevenrm #1}}$}{#2}}}

\font\twelvebf=cmbx10 scaled\magstep 1
\font\twelverm=cmr10 scaled\magstep 1
\font\twelveit=cmti10 scaled\magstep 1
\font\elevenbfit=cmbxti10 scaled\magstephalf
\font\elevenbf=cmbx10 scaled\magstephalf
\font\elevenrm=cmr10 scaled\magstephalf
\font\elevenit=cmti10 scaled\magstephalf
\font\bfit=cmbxti10
\font\tenbf=cmbx10
\font\tenrm=cmr10
\font\tenit=cmti10
\font\ninebf=cmbx9
\font\ninerm=cmr9
\font\nineit=cmti9
\font\eightbf=cmbx8
\font\eightrm=cmr8
\font\eightit=cmti8

\def\be{\begin{equation}}
\def\ee{\end{equation}}
\def\bea{\begin{eqnarray}}
\def\eea{\end{eqnarray}}

\centerline{\tenbf BLACK HOLES FROM BLUE SPECTRA}
\baselineskip=22pt
\vspace{0.8cm}
\centerline{\tenrm JAMES E. LIDSEY}
\baselineskip=13pt
\centerline{\tenit NASA/Fermilab Astrophysics Center, }
\baselineskip=12pt
\centerline{\tenit Fermi National Accelerator Laboratory, Batavia, IL 60510,
U.S.A.}
\vspace{0.3cm}
\vspace{0.3cm}
\centerline{\tenrm B. J. CARR \& J. H. GILBERT}
\baselineskip=13pt
\centerline{\tenit Astronomy Unit, School of Mathematical Sciences,}
\baselineskip=12pt
\centerline{\tenit Queen Mary \& Westfield, Mile End Road, London, E1 4NS,
U.K.}
\vspace{0.9cm}
\abstracts{Blue primordial power spectra with a spectral index $n>1$
can lead to a significant production of primordial black holes in the
very early Universe. The evaporation of these objects leads to a
number of observational consequences and a model independent upper
limit of $n\approx 1.4$. In some cases this limit is strengthened to
$n=1.3$. Such limits may be employed to define the boundary to the
region of parameter space consistent with generalized inflationary
predictions. [To appear in Proceedings of the CASE WESTERN CMB
WORKSHOP, April 22-24 1994. Figures available on request from
J.H.Gilbert@qmw.ac.uk]}

\vfil
\twelverm   
\baselineskip=14pt
\section{Introduction: Towards an Observational Test of Inflation}
The positive detection of anisotropic structure in the temperature
distribution of the Cosmic Microwave Background (CMB) radiation has
opened up the possibility that the predictions of the inflationary
scenario may be testable within the near future\cite{these,stein}.
During inflation the scale factor grows exponentially, whilst the
Hubble radius $H^{-1} \sim 10^{-23}$cm remains almost
constant. Consequently the physical wavelength of a quantum
fluctuation in the scalar or graviton field soon exceeds $H^{-1}$ and
its amplitude then becomes `frozen'.  Once inflation has ended,
however, $H^{-1}$ increases faster than the scale factor, so the
fluctuations eventually reenter the Hubble radius during the
radiation- or matter-dominated eras. The fluctuations that exit around
60 e-foldings or so before reheating reenter with physical wavelengths
in the astrophysically interesting range $1$ Mpc - $10^4$
Mpc. Fluctuations in the graviton degrees of freedom result in a
stochastic background of primordial gravitational waves that have an
amplitude at reentry given by $\delta_{\rm GW} \approx V/ m^4_{\rm
Pl}$, where $m_{\rm Pl}$ is the Planck mass.  Fluctuations in the
inflaton field provide the seeds for galaxy formation via
gravitational instability and their amplitude at reentry is $ \delta
\approx V^{3/2} / ( {m^3_{\rm Pl}}{|V'|})$.  Both scalar and tensor
fluctuations lead to CMB anisotropies.

Since $H$ decreases as the inflaton field rolls down the potential,
the amplitude of the scalar and tensor fluctuations is
scale-dependent. This variation is most conveniently parametrized in
terms of the spectral indices, $n_T$ and $n$, defined by $\delta_{\rm
GW}(M) \propto M^{-n_T/6}$ and $\delta (M) \propto M^{(1-n)/6}$, where
$M$ is the mass scale associated with the Hubble radius at the epoch
of reentry. In general these spectral indices are themselves functions
of scale. However, scales relevant to large-scale structure and CMB
experiments probe only 9 e-foldings or so of the inflationary
expansion, and since the scalar field must be rolling slowly for
inflation to occur in the first place, these scales correspond to a
very small portion of the inflationary potential. Hence, it is
reasonable to suppose that the spectral indices are indeed constant
over the scales of interest\cite{turner}.

When comparing theory with experiment, it is conventional to expand
the CMB temperature fluctuations on the sky in terms of spherical
harmonics. The angular correlation function predicted from theory is
then given by an average over all observer positions:
\be
\label{leg}
\langle \delta T (\theta ) \delta T (0) \rangle = \sum_{l=2}^{\infty}
\frac{2l+1}{4\pi} C_{l} P_l (\cos \theta ) ,
\ee
where the $P_l$'s are Legendre polynomials and a given multipole $l$
corresponds to an angular scale $\theta /1^o \approx 60/l$.  The
$C_l$'s corresponding to the scalar $(C_{l,S})$ and tensor $(C_{l,T})$
fluctuations are determined once the precise functional forms of the
spectra have been specified.  In the limit that $|n_T|$ and $|n-1|$
are constant and small, the ratio of the $l=2$ multipoles is uniquely
determined by the tensor spectral index:
\be
\label{consistency}
R\equiv \frac{C_{2,T}}{C_{2,S}} \approx -7 n_T .
\ee

This {\em consistency equation} is a fairly generic prediction of
inflation\cite{con}. The quantities in this expression are measurable,
at least in principle, and it forms the basis for an observational
test of the scenario. If no reionization occurs, experiments on scales
$\theta \ge 2^o$ ($l \le 30$) provide a measure of the sum of the
scalar and tensor contributions: $C_l =C_{l,S}+C_{l,T}$. Since the
gravitational waves do not produce a measurable contribution to the
CMB anisotropy for $\theta \le 2^o$, one might hope to determine
$C_{2,T}$ and $C_{2,S}$ separately from a combination of small and
large angle CMB experiments. A test of inflation would then require a
separate determination of the tensor spectral index. At present it
seems that this requires a direct detection of the gravitational wave
spectrum. Recent calculations\cite{direct} suggest that the maximum
present-day contribution per octave of the gravitational waves is
$\Omega_{\rm GW}h^2 \le 7\times 10^{-15}$. This is too weak to be
detectable by the Laser Interferometer Gravity-Wave Observatories,
although the proposed beam-in-space experiment has a peak sensitivity
of $\Omega_{\rm GW} \approx 10^{-16}$ at $10^{-4}$ Hz and may be able
to detect such a background.

\section{Constraining the Scalar Spectral Index with PBHs}

The scalar spectral index is much easier to measure and we would like
to have an expression equivalent to the consistency equation that
relates the $C_2$'s to $n$. In general, the relationship between $n_T$
and $n$ is model dependent, but for the special case of an exponential
potential, we have $n-1=n_T$. Hence, in the parameter space $(R,n)$,
this model corresponds to the line $R=7(1-n)$.  It is important to
note, however, that other families of potentials will lead to
different trajectories in this space. For example, any potential of
the form $V=V_0[1 \pm 2\pi |n-1| \phi^2 /m^2_{\rm Pl} ]$ leads to a
measurable deviation of $n$ from unity, whilst predicting that
$R\approx 0$.\cite{me} One could imagine taking all the current
candidates for the inflationary potential and predicting their
separate trajectories in this parameter space\cite{stein}. A
superposition of these paths would then define a target that
represented a generalized prediction of inflation in some sense. The
problem with this approach, however, would be in deciding which models
should be included in the target. Whilst a given model may appear to
be natural to one person, someone else may deem it unnatural.

It seems to us that this subjective element in the testing procedure
must be eliminated if such an approach is to be developed further. To
accomplish this it is necessary to search for model-independent
constraints on the parameters $R$ and $n$. Recently an upper limit on
the spectral index was derived from considering the formation and
subsequent evaporation of primordial black holes
(PBHs)\cite{CGL}. This limit is independent of any gravitational wave
contribution to the CMB anisotropy and therefore defines a boundary to
the observational target of inflationary predictions in $(R,n)$ space.

The idea behind the argument is rather simple. During inflation the
first scales to leave the Hubble radius are the last to come back in
and this implies that the very last fluctuation to leave is the first
to return. In the simplest case, the fluctuation on this scale will be
spherically symmetric and Gaussian distributed with an rms amplitude
given by $\delta (t_{e})$, where $t_{e} \sim H^{-1} $ is the time when
inflation ends. In some regions of the post-inflationary Universe, the
fluctuation will be sufficiently large that the collapse of a local
region into a black hole will become inevitable. The higher the rms
amplitude the higher the fraction of the Universe forming PBHs. The
observational consequences of the evaporation of these black holes
then leads to upper limits on the number that may form and hence on
the magnitude of $\delta (t_{e})$. An upper limit on $n$ is therefore
derived by normalizing the power spectrum on the quadrupole scale,
$M_Q \sim 10^{57}$g, and assuming that the spectral index is constant.

PBHs are never produced in sufficient numbers to be interesting if
$n<1$, but they could be if the spectrum is `blue' with $n>1$. The
precise form of the constraints depends crucially on how the Universe
is reheated, however, and we now proceed to discuss the separate cases
of efficient and inefficient reheating.

\subsection{Efficient Reheating}

When an overdense region with equation of state $p=\gamma \rho$ stops
expanding, it must have a size greater than $\sqrt{\gamma}$ times the
horizon size in order to collapse against the pressure and this
requires that $\delta (t_{e}) > \gamma$. It follows that the
probability of a region of mass $M$ forming a PBH is\cite{C1975}
\be
\label{beta0}
\beta_0 (M) \approx \delta (M) \exp \left( -\frac{\gamma^2}{2\delta^2
(M)} \right) .
\ee
Because of the exponential expansion PBHs that form before or during
inflation have no observational consequences.  Furthermore, if the
reheating process is very efficient, the false vacuum energy is
rapidly converted into relativistic particles with a reheating
temperature $T_{\rm RH} /T_{\rm Pl} = (t_e/t_{\rm Pl} )^{-1/2}$. The
mass of a PBH forming at this time is $M_{\rm RH} /m_{\rm Pl} \approx
t_{\rm RH} / t_{\rm Pl}$ and once it has formed, a PBH of this mass
will evaporate at a time $t_{\rm evap} \approx (M_{\rm RH} /m_{\rm
Pl})^3 t_{\rm Pl}$. Eq. (\ref{beta0}) then implies that for a blue
spectrum, $\beta_0(M)$ decreases exponentially for $M>M_{\rm RH}$, so
we may regard the PBH mass spectrum as effectively being a
$\delta$-function at $M_{\rm RH}$.

The constraints on $\beta_0 (M)$ in the range $10^{10}{\rm g} \le M\le
10^{17}{\rm g}$ were recently summarized\cite{CL}.  In particular,
PBHs with an initial mass $\sim 10^{15}$g would evaporate at the
present epoch and may contribute appreciably to the observed gamma-ray
and cosmic-ray spectra at 100 MeV. On the other hand, $10^{10} $g PBHs
have a lifetime $\sim$ 1 sec and, if produced in sufficient numbers,
their evaporations would lead to the photodissociation of deuterium
immediately after the nuclesynthesis era. PBHs of mass slightly below
$10^{10}$g could alter the baryon-to-photon ratio just prior to
nucleosynthesis. In our paper\cite{CGL} we consider the constraints on
$\beta_0 (M)$ below $10^{10}$g. In this region there is a potentially
stronger constraint on the spectral index if evaporating PBHs leave
stable Planck mass relics\cite{BCL}. Although the formation of such
objects has not been proved conclusively, it would be surprising if
quantum gravity effects did not become important once the PBH had
evaporated down to the Planck mass and various arguments have been
developed in the literature suggesting the formation of such objects
is likely\cite{CGL}.

To derive the observational constraint from PBH relics one proceeds as
follows: a lower limit on $M_{\rm RH}$ is derived by assuming that the
observed quadrupole anisotropy is due entirely to gravitational
waves. This implies\cite{direct} that the expansion rate of the
Universe during the last 60 e-foldings of inflation cannot exceed $3
\times 10^{-5}m_{\rm Pl}$ and leads to an upper limit on the reheat
temperature of $\sim 10^{16}$ GeV. This corresponds to a minimum mass
$\sim 1$g. The observational constraint from the relics derives from
the fact that they cannot have more than the critical density at the
present epoch, i.e. $\Omega_{\rm rel} <1$. The precise form of the
constraint depends on whether the evaporating PBHs dominate the energy
density of the Universe before they evaporate. Since the ratio of PBH
density to radiation density increases as $ {t}^{1/2}$, the PBHs do
not dominate at evaporation if $\beta_0 < ( M_{\rm RH} / m_{\rm
Pl})^{-1} $. If this condition is satisfied, the constraint that
$\Omega_{\rm rel}$ does not exceed unity becomes
\be
\label{relic}
\beta_0 (M) < 10^{-27} \left( \frac{M}{m_{\rm Pl}} \right)^{3/2} .
\ee
If, on the other hand, the PBHs dominate the density at evaporation,
most of the background photons derive from the PBHs and the constraint
becomes $M>10^6$g. In other words, only relics formed from PBHs
smaller than $ 10^6$g can contribute significantly to the current
energy density and above this critical mass the entropy constraint
takes over.

We have completed a detailed analysis and derived the constraints on
$\delta (M_{\rm RH})$ for all mass scales above $1$g. These results
are illustrated in Figure 1. It is seen that the strongest limit on
the spectral index derives from the relic constraint and by combining
Eqs. (\ref{beta0}) and (\ref{relic}) one finds that
\be
\label{deltarelic}
\delta (M_{\rm RH}) < 0.13 \left[ 17 -{\rm log}_{10} \left(
\frac{M_{\rm RH}} {m_{\rm Pl}} \right) \right]^{-1/2} .
\ee
If we normalize on the quadrupole scale, $M_Q$, the rms amplitude on a
smaller scale $M$ is $\delta (M) =\delta_Q (M/M_Q)^{(1-n)/6}$, where
$\delta_Q \approx 3.8 \times 10^{-6}$. Hence, we conclude from
Eq. (\ref{deltarelic}) that the limit on the spectral index is $n\le
1.4$ for $M_{\rm RH} \approx 1{\rm g}$, corresponding to a reheat
temperature $\sim 10^{16} $GeV, and $n\le 1.5$ for $M_{\rm RH} \approx
10^6$g, corresponding to $T_{\rm RH} \sim 10^{14} $GeV. The
constraints associated with higher masses (i.e. lower reheat
temperatures) are calculated by a similar procedure and are summarized
in Figure 2.

\vspace{4.7in}

 \fcaption{The constraints on the rms amplitude of the scalar
fluctuation spectrum immediately after inflation if the equation of
state is radiation-like ($\gamma =1/3$).  The origin of the
constraints above $10^{10}$g is summarized by Carr and
Lidsey\cite{CL}. If PBHs do not leave behind stable relics after
evaporation, the strongest upper bound on the spectral index is given
by the dashed line which joins the COBE/DMR point and the deuterium
constraint at $10^{10}$g. This limit applies for reheat temperatures
$\sim 10^9 $GeV. If relics are formed, the limit is strengthened at
higher reheat temperatures as indicated.}

\newpage

\subsection{Inefficient Reheating}

Thus far we have assumed that the reheating of the Universe to
relativistic particles occurs on a timescale much less than
$H^{-1}$. However, when inflation ends by means of a second-order
phase transition, the scalar field undergoes coherent oscillations in
the potential minimum from the time $t_1 \sim H^{-1}$ until it decays
at a time $t_2 \sim \Gamma^{-1}$, where $\Gamma$ is its decay
width. During this interval, the Universe is effectively dominated by
a dust-like fluid\cite{dust} with $\gamma =0$, and Eq. (\ref{beta0})
does not apply.  PBHs will still form in this case, but the fraction
of the Universe going into PBHs is now determined by the probability
that regions are sufficiently spherically symmetric to collapse within
their Schwarzschild radius. This fraction is given by\cite{pol}
\be
\label{beta}
\beta (M) \approx 2\times 10^{-2} [\delta (M) ]^{13/2}
\ee
and the observational constraints on the probability of PBH formation
are altered because they now have an extended mass spectrum. The range
over which this spectrum applies is defined by $M_1\le M\le M_{\rm
max}$, where $M_1$ is the horizon mass immediately after inflation and
$M_{\rm max}$ is the mass of a configuration that just detaches itself
from the universal expansion at $t_2$. It is determined implicitly
by\cite{pol}
\be
\label{masscondition}
M_{\rm max} = [\delta (M_{\rm max}) ]^{3/2} \left( \frac{t_2}{t_{\rm
Pl}} \right) m_{\rm Pl} .
\ee
The constraints on $\beta (M)$ are related to the associated
constraints on $\beta_0 (M)$ via the relation
\be
\label{beta0beta}
\beta (M) =\beta_0 (M) \left( \frac{t_2}{t_{\rm Pl}} \right)^{1/2}
\left( \frac{M}{m_{\rm Pl}} \right)^{-1/2} .
\ee
This is a useful expression because it implies that the limits on PBH
formation during the dust phase can be calculated directly from the
constraints which apply if there is no dust phase. We first assume
$t_1$ is fixed and vary the epoch $t_2 = \Gamma^{-1}$ at which the
dust era ends. For a given value of $t_2$, the mass range of PBHs
forming during the dust phase goes from $M_1$ to the mass given by
Eq. (\ref{masscondition}) and for each value of $M$ the constraint on
$\beta (M)$ is given by Eq. (\ref{beta0beta}). The corresponding limit
on $\delta (M)$ follows from Eq. (\ref{beta}) and the limit on $n$ is
derived by normalizing on the quadrupole scale as before.  The reheat
temperature is $T_{\rm RH} \approx (\Gamma t_{\rm Pl})^{1/2} m_{\rm
Pl}$ and an upper limit on $\Gamma t_{\rm Pl}$ follows from the
requirement that baryogenesis must proceed after reheating. It is
generally accepted that the lowest temperature for which the observed
baryon asymmetry may be generated is the electroweak scale, $\sim 10^3
$GeV, corresponding to $\Gamma t_{\rm Pl} \sim 10^{-30}$.

The new constraints on the spectral index are also shown in Figure
2. For a relatively short dust phase, only the relic limit will be
altered, since PBHs above $10^{10}$g will not form during the dust
phase. For lower reheat temperatures, however, more massive PBHs form
and provide the strongest constraint. The deuterium constraint applies
for $10^{-17} \ge \Gamma t_{\rm Pl} \ge 10^{-23}$ and the gamma-ray
limit for $10^{-23}\ge \Gamma t_{\rm Pl} \ge 10^{-30}$. From this
figure we arrive at an upper limit of $n=1.4$ if PBHs form relics and
there is a dust-phase after inflation.

\phantom{blabla}

\vspace{4.5in}

\fcaption{Illustrating the constraints on the spectral index
arising from the overproduction of primordial black holes, the shaded
area being excluded. The lower line applies if there is a dust phase
immediately after inflation, in which case the ordinate is $ {\rm
log}_{10} \Gamma t_{\rm Pl}$, the upper line if there is no dust phase
in which case it is ${\rm log}_{10}(t_{\rm Pl}/t_e)$. The constraints
depend on the reheat temperature $T_{\rm RH} \approx 10^{18} (\Gamma
t_{\rm Pl} )^{1/2}$GeV, where $\Gamma$ is the decay width of the
scalar field that decays into relativistic particles. The
$n$-independent upper and lower limits on the decay width arise from
assuming the COBE/DMR detection is due entirely to gravitational waves
(shaded line) and from requiring that baryogenesis can only proceed
above the electroweak scale (dashed line). The dotted horizontal line
indicates the CMB distortion limit of Hu et al \cite{FIRAS}. For
reheat temperatures above $T_{\rm RH} \approx 10^{9.5}$GeV the most
important constraint arises from the requirement that any Planck mass
relics left over from the final stages of PBH evaporation should have
less than the critical density at the present epoch. For lower reheat
temperatures, more massive PBHs may form and the strongest constraints
then arise from the photodissociation of deuterium by evaporating
$10^{10}$g PBHs, from the observed gamma-ray background in the energy
range $0.1-1$ GeV, or from the distortions of the CMB.}

\section{Conclusion}

One criticism of these limits is that they assume the spectral index
is constant over the full range of scales. To answer this point we
note that the general inflationary potential leading to spectra with
{\em constant} spectral index $n>1$ is a combination of trigonometric
functions with a Taylor expansion of the form\cite{CL}
\be
\label{potential}
V=V_0\left[1 + 2\pi (n-1) \frac{\phi^2}{ m^2_{\rm Pl}} \right] .
\ee
Any potential that leads to $n>1$ will have a Taylor expansion of this
form. It follows that, as the field rolls towards the minimum, the
approximation of Eq. (\ref{potential}) to the general trigonometric
potential becomes more accurate and so the variations in the spectral
index become smaller. Consequently, one need only show that the
spectral index is effectively constant over the scales corresponding
to large-scale structure (53-60 e-foldings from the end of
inflation). It is straightforward to show that this is the
case\cite{CGL}.

We conclude that the formation of PBHs from quantum vacuum
fluctuations immediately after inflation constrains the scalar
spectral index to be less than $1.4$. It is important to emphasize
that, because our limit spans the large range of scales from $\sim 1$g
to $\sim 10^{57}$g, (for comparison, all large-scale structure
measurements span only 10 decades of scale), it is essentially
independent of the errors that arise in the COBE normalization from
possible tensor contributions, cosmic variance and the effect of the
Doppler peak on the low multipole anisotropies. Moreover, since we
have normalized on COBE scales, the limit is also independent of the
precise form of dark matter and hence the bias parameter. In effect,
once the COBE normalization is specified, the limit is independent of
the cosmological model, although it does assume that $\Omega_0=1$, as
predicted by most inflationary models. Therefore this limit can be
employed to define a boundary to the target of inflationary
predictions in $(R,n)$ space.

\section{Acknowledgements}

JHG and JEL are supported by the Science and Engineering Research
Council (SERC), UK. JEL is supported at Fermilab by the DOE and NASA
under Grant No. NAGW-2381. We would like to thank R. C. Caldwell,
A. R. Liddle, K. Maeda, A. Polnarev and D. Scott for fruitful
discussions about this work. JEL thanks L. M. Krauss for organizing a
very enjoyable and informative workshop.

\end{document}